\documentclass[12pt]{article}
\usepackage{amsfonts,amssymb,epsfig,amsmath}
\usepackage{color}

\renewcommand{\baselinestretch}{1.2}

\newcommand{\be}{\begin{eqnarray}}
\newcommand{\ee}{\end{eqnarray}}

\newcommand{\bn}{\begin{enumerate}}
\newcommand{\en}{\end{enumerate}}

\begin{document}

\makeatletter \@addtoreset{equation}{section} \makeatother
\renewcommand{\theequation}{\thesection.\arabic{equation}}
\renewcommand{\thefootnote}{\alph{footnote}}

\begin{titlepage}

\begin{center}

\vspace{2cm}

{\Large\bf Fractional Quantum Hall Effect and M-Theory}

\vspace{2cm}

\renewcommand{\thefootnote}{\alph{footnote}}

{\large 
Cumrun Vafa}

\vspace{0.7cm}

\textit{ Jefferson Physical Laboratory, Harvard University, Cambridge,
MA 02138, USA.}\\

\vspace{0.7cm}

\end{center}

\vspace{1cm}

\begin{abstract}

We propose a unifying model for FQHE which on the one hand connects it to recent developments in string theory
and on the other hand leads to new predictions for the principal series of experimentally observed FQH
systems with filling fraction $\nu ={n\over 2n\pm1}$
as well as those with $\nu ={m\over m+2}$.  Our model relates these series to minimal unitary models of the Virasoro
and superVirasoro algebra and is based on $SL(2, {\bf C})$ Chern-Simons theory in Euclidean
space or $SL(2,{\bf R})\times SL(2,{\bf R})$ Chern-Simons theory in Minkowski space.  This theory,
which has also been proposed as a soluble model for 2+1 dimensional quantum gravity,
and its N=1 supersymmetric cousin, provide effective descriptions
of FQHE.   The principal series corresponds to quantized levels for the two
$SL(2,{\bf R})$'s such that the diagonal $SL(2,{\bf R})$ has level $ 1$.
The model predicts, contrary to standard lore, that for principal series of FQH systems the quasiholes possess non-abelian statistics.
For the multi-layer case we propose that complex ADE Chern-Simons theories provide effective
descriptions, where the rank of the ADE is mapped to the number of layers.   
Six dimensional $(2,0)$ ADE theories on the Riemann surface $\Sigma$ provides a realization of FQH
systems  in M-theory.   Moreover we propose that the q-deformed version of Chern-Simons theories are related
to the anisotropic limit of FQH systems which splits the zeroes of the Laughlin wave function.
Extensions of the model to 3+1 dimensions, which realize topological insulators
with non-abelian topologically twisted Yang-Mills theory is pointed out.

\end{abstract}

\end{titlepage}

\renewcommand{\thefootnote}{\arabic{footnote}}

\setcounter{footnote}{0}

\renewcommand{\baselinestretch}{1}

\tableofcontents

\renewcommand{\baselinestretch}{1.2}
\section{Introduction}
Since its discovery \cite{tsui} fractional quantum Hall effect has attracted a great deal of attention by both theorists and experimentalists.
On the theory side, with the proposal of Laughlin \cite{Laughlin:1983fy} as well as development of other theoretical
ideas such as hierarchy states of Haldane and Halperin \cite{haldane,halperin}, and Jain's composite fermion theory \cite{Jain:1989tx},
many of the observed filling fractions were explained, including the prediction of abelian anyonic statistics.  Moreover, non-abelian
statistics which was anticipated in \cite{Fro} (see also \cite{ver}), was connected to FQHE in \cite{wen,Moore:1991ks} and further extensions were considered \cite{Read:1998ed} (see also \cite{Wen:2008oda}).  These constructions utilize Chern-Simons theory based on Witten's discovery of
non-abelian braiding in these theories \cite{Witten:1988hf}. 
To date neither the abelian nor the non-abelian statistics has been fully verified experimentally.

In this brief note we propose a new model for FQHE which connects it on the one hand to recent developments in string theory
and on the other hand leads to new predictions for the principal series of FQHE's with filling fraction $\nu ={n\over 2n\pm1}$, as well
as those with $\nu={m\over m+2}$.
In particular our model predicts that for principal series of FQHE's, unlike the prediction of hierarchy states and composite fermion theory, the quasiholes possess non-abelian statistics, related
to Fusion algebra of $(2n,2n\pm1)$ unitary CFT minimal models.   For the filling fraction $\nu ={m\over m+2}$ we obtain the fusion algebra of SCFT unitary minimal models $(m,m+2)$.
Moreover the first in the CFT series correspond to Laughlin's $\nu ={1\over3}$ and the first in the SCFT series to Moore-Read state $\nu={1\over 2}$.  The
higher values of $n$ can be obtained by composite fermion model or hierarchy model and the ones with higher $m$ can be obtained
from Read-Rezayi states.  However, our models predict a different fusion algebra, and thus a different statistics than expected from the corresponding
constructions.\footnote{Except there are subtletlies for the fusion algerba
coefficents of Degenerate fields of Liouville theory, which we are currently studying \cite{tv}.}   In particular if our model is correct it would predict that essentially all the observed cases of FQHE involve non-abelian statistics,
which is potentially a welcome news for quantum computing \cite{Kitaev:1997wr}!  For a nice review of constructions of non-abelian
statistics in FQHE and connection to quantum computing see \cite{Nayak:2008zza}.  Moreover our model predicts, on a sharp edge,
charged downstream and nuetral upstream currents which distinguishes it for filling fraction $\nu=n/(2n+1)$ from the standard model of  FQH systems which predicts no upstream neutral currents.  Moreover we predict the Hall conductivity for these edge modes which is different from the hierarchy
or composite models.
Compared to the usual constructions ours has the advantage of having essentially no adjustable parameters and for the single
layer FQHE the assumption of unitarity picks these sequences in
our construction.

Even though there is a relation between the fusion algebra of the quasi-holes and the minimal models, it is important to note
that the relation to a CFT arises as descriptions on the 1+1 dimensional boundary of the sample, which is
echoed in the bulk by the associated monodromy structure of excitations in the 2+1 dimensional bulk. There are different boundary conditions that we can have in our model.  In one boundary condition,
which we identify as a superconducting interface, we obtain the minimal model chiral blocks
as effective description of the 1+1 edge modes which propagate only in one direction.  Another boundary condition, 
which is the more standard one corresponding to a sharp edge, we get a different
CFT which has the same block structure as the minimal model, but which leads to downstream charged currents
and neutral upstream currents.
The bulk theory does not depend on the choice of the boundary condition
and in either case is given by Chern-Simons theory based on complex gauge group $SL(2,{\bf C})$
studied in \cite{witcs} (see also \cite{Rocek:1985bk,Achucarro:1987vz} and related work \cite{Witten:1988hc}), and its
supersymmetric cousin (for recent discussions of $SL(2,{\bf C})$ Chern-Simons
theory see
\cite{Dimofte:2011jd,Gukov:2015sna}).  More precisely,  $SL(2,{\bf C})$ has a pair of levels $l=(k,\sigma)$, where $k$ is an integer
(specifying the level of $SU(2)\subset SL(2,{\bf C})$) and $\sigma$ is a real parameter\footnote{This is related
to cosmological constant in the gravitational picture of the theory.}.  For $l=(\pm 1,4n\pm 1)$ we 
obtain the principal series with filling fractions $\nu={n\over {2n\pm 1}}$.
 In the Minkowski signature this corresponds to $SL(2,{\bf R})\times SL(2,{\bf R})$ Chern-Simons theory
and for $|k|=1$ the values for $\sigma$ which yield the principal series of filling fractions are {\it exactly} the ones
which would follow if we make the individual $SO(2)$'s in the $SL(2, {\bf R})$'s quantized with levels
$(-2n, 2n\pm1)$.  It is natural to conjecture that the higher Jain series with filling fractions $\nu ={n\over 
2nk'\pm1}$ correspond in our setup to the two $SL(2,{\bf R})$ levels being given by $(-2n,2nk'\pm 1)$,
which would correspond to the level $k=2n(k'-1)\pm1$ for the diagonal $SL(2,R)$.

In  \cite{Witten:1988hc} (p.75-76)  Witten specifically suggests viewing
minimal Virasoro models in 2d as holographic realization of 2+1 gravity.  In our context this would
suggest that the FQHE is holographically encoding 3d gravity and large enough Laughlin quasi-holes are actually black-holes!  In this context the $1+1$ dimensional
edge theory on the boundary of the sample is holographic dual to the FQHE bulk interpreted
as a gravitational theory.  In other words we can realize
holography in the lab!
It is natural to connect our model
to the observations in \cite{Haldane:2011ia} involving an emergent geometry in FQHE.  
Indeed the elements of Haldane's proposal, and in particular the appearance of $SL(2,{\bf R})\times SL(2,{\bf R})$ in his setup are in harmony with the picture proposed here \cite{DGV}.

These constructions were motivated by string theory which in turn leads to a proposal for
the corresponding Hamiltonian.   These involve compactification of 6d $(2,0)$ theories on a surface $\big[\Sigma \times {\bf R}\big] \times S^3_{k,{b^2}}$,
and $\Sigma$ is identified with the plane of the FQHE and ${\bf R}$ with time, $S^3_k$ is the lens space $S^3/{\bf Z_k}$ and $b^2$ is a squashing
parameter\footnote{We will be interested in the limit where $b^2$ is a negative rational number, so
this is defined in the sense of analytic continuation.}  for $S_k^3$.  There is an ADE classification for 6d $(2,0)$ theories and we identify
the rank of the ADE with the number of layers in the FQHE.   This leads to our {\it  identification of
the effective theory of FQH systems with complex} ADE {\it  Chern-Simons theories with the rank of} ADE {\it
corresponding to the number of layers}.
The $A_1$ case, correponding to single layer leads in Euclidean description, to  $SL(2,{\bf C})$ Chern-Simons effective theory on $\Sigma \times {\bf R}$.

 The organization of this paper is as follows:  In section 2 we explain the heuristic
motivation for this model and  the connection with Liouville theory and 2d minimal CFT's.  Also discussed there is the connection with
non-abelian statistics for quasi-holes.  In section 3 we propose a Hamiltonian whose improved Berry's connection is expected to yield the results outlined
in section 2 for the monodromy of the quasihole in Liouville theory.  In section 4 we sharpen our proposal by embedding
it in string theory.  Section 4.2
is a self-contained summary of the model and readers who are not interested in the motivation for our model
can go directly to that section.   The string theory perspective leads to reformulation of the effective
 theory of the (single layer) FQH systems in the bulk as complex $SL(2,{\bf C})$ Chern-Simons theory
and provides various generalizations of it, including predictions of new defects for FQHE, as well as potential applications to multi-layer and anisotrpic FQH systems. We also explain the heterotic aspect of the edge states for our model as well as the different choices of boundary conditions.   Moreover we comment on the possibility of lifting the construction
to one higher dimension and potentially exciting realization of topological insulators
with topologically twisted non-abelian gauge symmetry in 3+1 dimensions in the lab!

\section{Basic idea}
Let us start with the Laughlin wave function  \cite{Laughlin:1983fy} for electrons at positions $z_i$ and quasi-holes at positions $\zeta_a$ with filling fraction $\nu =1/m$, given by
$$\psi (z_i,\zeta_a)=\prod_{i,a} (z_i-\zeta_a) \prod_{i<j} {(z_i-z_j)}^{1\over \nu}{\rm exp}(-B\sum_i |z_i|^2)$$
This has proven to be a powerful model for FQHE (for a beautiful introduction to this subject see \cite{Wen}) .   This problem has been mapped to the study of RCFT's \cite{Moore:1991ks},  whose basic blocks satisfy the Verlinde algebra \cite{Verlinde:1988sn}, and which has been elegantly systematized in \cite{Moore:1988qv} even though there is no full classification.
  Consider $c=1$ theory
at radius $R^2=\nu$, and consider the chiral vertex operators $V(z_i)={\rm exp}(i \phi(z_i)/\sqrt \nu)$, and $W(\zeta_a)={\rm exp}(i \sqrt{\nu} \phi (\zeta_a))$.
Then the holomorphic part of the wave function (i.e. dropping the B-field part) can be captured (up to prefactors depending only on $\zeta_a$) by
$$\psi (z_i, \zeta_a) =\langle \prod_{i,a} V(z_i) W(\zeta_a)\rangle$$
Moreover to compute physical amplitudes one considers
$$\langle {\cal O}\rangle =\int d^2z_i \big[ \psi^*(z_i,\zeta_a) \ {\cal O} \ \psi (z_i,\zeta_a)\big]$$
One can also add a chemical potential $\mu$ for the fermions and add to the action
$\mu \int d^2z\  e^{i\phi/{\sqrt \nu}}$
for which the above term corresponds to the term $\mu^N$ where $N$ is the number of electrons.
Moreover one imagines $B$ field as being given by an additional smeared term by adding to the above ${\rm exp}\int B\phi$.  
However already there is a clash
with conformal field theory paradigm:  In conformal theories we usually do not integrate over the position of fields unless they have dimension $1$.  In the condensed matter context we are discussing, this is not
strictly necessary (see \cite{read} and references therein)\footnote{We thank N. Read for discussions on this point.} when we have a B-field.  However, in the absence
of B-field, as in studying superconductor phases where
we can ignore the B-field, the dimensions should be $1$.  To make this natural for FQH system
imagine a thin strip of material which on both
sides is in contact with a superconducting material.  
With this in mind, it would be interesting to see what demanding marginality of this operator would imply about the possible bulk theories.
Our strategy would be to first study implications that this would have in identifying the bulk theory,
and then using other boundary conditions such as the sharp edge one which would be interesting
for charged edge currents.  Later in section 3.3 we return to the question of introducing back the magnetic field and show how it can be incorporated in our setup.

Demanding that  ${\rm exp}(i \phi/\sqrt \nu)$  have dimension 1 is rather significant.  In the $c=1$ model it has dimension $h={1\over 2 \nu}\not= 1$.  So how can this be rescued?  
It is natural to add a background term to the action $Q'R\phi$ where $R$ is the curvature (which disappers in
flat space), in order to make the dimension of this field 1,
without affecting the above realization of the wave function.  
In such a case a vertex operator ${\rm exp}( \alpha \phi)$ will have dimension 
$$h_\alpha ={-1\over 2}(\alpha( \alpha +Q'))$$
To make the dimension of  ${\rm exp}(i \phi/\sqrt \nu)$ equal to 1, we need to take\footnote{We have chosen an unconventional normalization of Liouville field in 
order to make the dimension of the fields match the more familiar normalization for $c=1$ theory.}
$${Q'\over \sqrt 2}=Q={1\over b}+b=i(\sqrt{2\nu}-{1\over {\sqrt {2 \nu}}})$$
where
$$b={-i\over \sqrt {2 \nu}}.$$
Moreover the central charge of the 2d chiral theory is
$$c=1+6 Q^2=1-3 {({2 \nu} -1)^2\over \nu }$$
If we put all the ingredients together, we get the Liouville theory  (see \cite{Teschner:2001rv}  and lectures in \cite{Zamolodchikov} for a review of Liouville theory)
$$S=\int d^2z\big[ \ {1\over 8\pi}\partial \phi {\overline \partial} \phi +iQ' R \phi +\mu\  {\rm exp}(i\phi /{\sqrt \nu}) \big]$$
Related ideas have recently been suggested independently in \cite{Ferrari:2014yba,Can:2014tca,Klevtsov:2015eda}, but with a different motivation and without fixing the relation between $Q$ and $\nu$ and which reach rather different conclusions from the present paper\footnote{We thank A. Abanov for pointing out these papers to us.}.
So far we have assumed that $\psi(z_i)$ is a single valued wave function which in Laughlin's case is related
to $1/\nu =m \in {\bf Z}$. 
 This is because electron's wave function should be single valued.  But now we overcome the condition that $1/\nu \in {\bf Z}$ and propose the Liouville theory for all fractional values of $\nu$ as giving us
the overlaps $\langle \psi | \psi \rangle$ (i.e. the overlap of the multi-particle wave function ignoring the B-field term).  It is known in the context of Liouville that one can associate chiral blocks
to the wave functions \cite{Dotsenko:1984nm}.  This gives us a prescription as to how to compute
overlaps of multi-valued wave functions. Roughly speaking we can view them as wave function for suitable excitations which can
have non-trivial statistics, and so in particular the wave functions do not have to be single valued.  However this is not precise, because extracting
physical quantities from these wave functions is more subtle than the usual prescription (as we review in the next section) and is not simply
given by $\int d^2z \ \psi^* {\cal O} \psi$.     In section 3 we will explain how this multi-valued wave function is
related to a microscopic theory where we have well defined single valued wave functions.

Relaxing the assumption that $1/\nu$ is integer, we 
consider FQHE with filling fraction $\nu ={n\over m}$.  If we evaluate the central charge of the Liouville theory for that case we get
$$c=1-6{(2n-m)^2\over 2nm}$$
which is the same as central charge of 2d CFT minimal model $(2n,m)$ \cite{Belavin:1984vu}.  Note that $m$ needs to be odd in this context
as otherwise $2n$ and $m$ are not relatively prime!  Moreover Liouville theory has a special set of fields (called `degenerate fields') whose OPE  \cite{Cremmer:1993qf}
realize the operator algebra for the minimal model when restricted to $-b^2$ a rational number\footnote{We thank
Joerg Teschner for a discussion on this point.}.  Namely consider
$$\Phi_{r,s}={\rm exp}\big[ i (r-1){\sqrt \nu}\phi +(s-1){1\over 2\sqrt \nu}\big]$$
for $1\leq r< 2n, 1\leq s < m$. Then these degenerate fields of Liouville
realize the operator algebra of the $(2n,m)$ minimal models (see \cite{Ginsparg:1988ui} for a review)\footnote{There are subtleties in this statement which arises because
of analytic continuation of Liouville theory from $b^2>0$ to $b^2<0$.  See in particular 
\cite{Zamolodchikov:2005fy,Ribault:2015sxa,Harlow:2011ny}.}
$$\Phi_{r_1,s_1} \times \Phi_{r_2,s_2}=\sum_{\substack{k=1+|r_1-r_2|, k+r_1+r_2+1=0\ {\rm mod}\ 2\\ l=1+|s_1-s_2|, l+s_1+s_2+1=0\ {\rm mod}\ 2}}^{\substack{k=r_1+r_2-1\\ l=s_1+s_2-1}}\Phi_{k,l}$$
Strictly speaking the above discussion is in the context of a superconducting boundary
condition where the $B$-field is absent and the above algebra reflects
the algebra of neutral currents in an interface with a superconductor.   We use
this to identify what the bulk theory is (i.e. Chern-Simons theory based on $SL(2,{\bf C})$ guage group).  We postpone this discussion to section 4, where we connnect
it to string theoretic motivations for identifying the bulk.  As we discuss there, the same
bulk theory with a {\it different} boundary condition leads
to FQH system with a sharp edge boundary, where the $\Phi_{r,s}$ blocks are related
to the downstream electric currents in the $s$ label and neutral upstream currents labeled by $r$.  In particular if one considers $\Phi_{1,s}$, these
correspond to the usual $s-1$ quasi-hole states, which correspond to insertion of $\prod (z_i-\zeta)^{s-1}$ in the wave function, and form
a closed operator algebra generated by $\Phi_{1,2}$.
Moreover, given that the quasi-hole operators correspond to these operators,
we deduce that quasi-holes will possess the same non-abelian braiding properties as that of $(2n,m)$ models.
This is different from the hierarchy state or composite fermion model construction of these wave functions which predicts abelian statistics for quasiholes.  In particular taking the minimal quasi-hole $\Phi_{1,2}$ around another one (i.e. a $2\pi$ rotation) is expected to lead, in our model, to two dimensional fusion channel and the two phases one picks up
are given by the formulas for dimensions of blocks
$$\big[e^{2\pi i (h_{3,1}-2 h_{2,1})}, e^{2\pi i (h_{1,1}-2h_{2,1})}\big]=\big[e^{2\pi i\nu}, e^{-2\pi i (3\nu)}\big]=
\big[ e^{2\pi i {n\over m}}, e^{-6\pi i {n\over m}}\big].$$

We can now further restrict the choices of $m$:
Since the edge modes in the FQHE should have correlations which fall off rather than
grow with distance we would need
to restrict to theories which are unitary (see in particular the discussion in \cite{read}).
Applying this to when the FQH system is in an interface with superconductor, i.e. when
the edge theory is given by the Liouville theory, we should get
 unitary 2d CFT's with central charge less than 1 because
$b^2<0$, which fixes $m=2n\pm 1$ and corresponds to $(2n,2n\pm1)$ minimal model and filling fractions
$$\nu ={n\over 2n\pm 1}$$
So we find that the 2d unitary minimal models map exactly to the two principal series of FQHE which
have been experimentally observed!   This is a remarkable check of our proposal which connects
representations of minimal unitary 2d CFT's to principal series already observed in FQHE!  For the original
application of minimal unitary 2d CFT to critical phenomena see \cite{Friedan:1983xq}.
This is what we will obtain with a superconducting boundary condition.  It is also natural
to ask what we get
with the sharp edge boundary condition 
as is usually considered in the context of FQH system.  This we will
postpone to section 4, after we connect our model to $SL(2,{\bf C})$ Chern-Simons theory.
There we will find that the sharp edge 1+1 dimensional theory has
a central charge given by $(c_L,c_R)=(3-{6\over 2n},3-{6\over 2n\pm1})$
and the blocks are mixed between left- and right-moving sectors.

The above series give us only the odd denominator filling fractions and it is natural to ask how one can obtain the even
denominator ones as well.  This turns out to have a natural answer:  We simply extend the Liouville theory to ${\cal N}=1$ supersymmetric Liouville theory (see e.g.
\cite{Nakayama:2004vk,Belavin:2007gz})
whose action is given by
$$S=\int d^2z\big[ \ {1\over 8\pi}\partial \phi {\overline \partial} \phi +{1\over 2\pi}\big(\psi {\overline \partial} \psi +{\overline \psi}\partial {\overline \psi}\big)+iQ R \phi +2i \mu b^2{\overline \psi}\psi e^{b \phi}+ 2\pi b^2 \mu^2 e^{2b\phi} \big]$$
where $Q=b+1/b$,
With central charge ${2\over 3} c=\hat c =1+2Q^2$.  The case which corresonds to $(m,n)$ SCFT minimal models is when $b^2=- {n\over m}$ with $n-m =0\ {\rm mod} \ 2$
and the unitary minimal series $(m,m+2)$ corresponds to $n=m+2$.  The ${\overline \psi} \psi e^{b\phi}$ term has dimension one.   Note that the chiral wave function this
leads to in the free field realization is
$$\Psi =\prod_{i<j} (z_i-z_j)^{n\over m} \ {\rm Pf}\big[{1\over z_i-z_j}\big]$$
where ${\rm Pf}$ is the Pfaffian.  The $N=1$ unitary series corresponds to
$$\Psi =\prod_{i<j} (z_i-z_j)^{m+2\over m} \ {\rm Pf}\big[{1\over z_i-z_j}\big]$$
where $m=2,3,...$.  The first  element of the series, $m=2$, corresponds to the Moore-Read wave function \cite{Moore:1991ks}.  The filling fractions we get for the unitary $N=1$ case are $\nu ={m\over m+2}$.
These values for filling fractions also arise in Read-Rezayi's construction \cite{Read:1998ed} which generalizes the
Moore-Read state, but these models  (except possibly for the $m=2$ case) are distinct from ours.
The bulk theory in this case is a supersymmetric version of $SL(2,{\bf C})$, and the superconducting
boundary conditions lead to supersymmetric Liouville theory.  As in the non-supersymmetric case, one
can also consider the sharp edge boundary condition in these cases as well.

So far the relation we have found is rather suggestive.  In the next section we propose a microscopic Hamiltonian motivated
from string theory (see section 4) whose Berry's
connection, as we vary the position of quasiholes, is known to lead to the
braiding properties for the Liouville amplitudes, in some limit.  It is natural to expect a cousin of it also exists which gives
the one for supersymmetric Liouville amplitudes.

\section{A microscopic description}
Here we give a Hamiltonian description which is relevant for the Liouville phase of the theory, i.e. the superconducting interface
where the $B$-field is absent.  We first start with a simple model hamiltonian for a single
particle and then move on to the many particle case.

\subsection{Hamiltonian construction from $W(z)$}
We first discuss a single particle toy model and then we generalize it to the case at hand.
Consider a holomorphic function $W(z)$.  To this we will associate a Hamiltonian for a particle which is a bi-spinor (i.e. has
$2\times 2=4$ internal degrees of freedom) on the $z$-plane as follows:
$$H={1\over 2}p^2 +{1\over 2}\big |\partial_z W(z)|^2 +\partial^2 W(z) \cdot \sigma^+\otimes \sigma^- +{\overline \partial}^2 {\overline W}({\overline z})\cdot \sigma^-\otimes \sigma^+$$
Then for each cricial point $\partial W=0$ we get a ground state for this theory, which turns out to have exactly zero energy (because this
system secretly enjoys 4 units of supersymmetry--see \cite{Cecotti:1990wz}).
 It can also be written as\footnote{In particular $H=Q^2$ for a $Q$ which the reader
can easily identify.}
$$H={1\over 2}p^2 +{1\over 2}\big |\partial_z W(z)|^2 +\big[ \partial^2 W(z)\cdot b^\dagger_Lb_R +{\overline \partial}^2 {\overline W}({\overline z})\cdot b^{\dagger}_R b_L\big]$$
where $b^\dagger_{L,R},b_{L,R}$ form a pair of fermionic creation/annihilation operators with the non-vanishing anti-commutations being
$$\{b_L,b_L^\dagger \}=1,\qquad \{b_R,b_R^{\dagger}\}=1$$
The non-abelian berry's connection \cite{{Berry:1984jv,Wilczek:1983cy}}
for the degenerate ground states of this
system as a function of parameters defining $W$ satisfies a beautiful set of equations known as the $tt^*$ geometry \cite{Cecotti:1991me}.
To compute this connection we need to compute overlap between the ground state wave functions $\langle \psi_i |\psi_j\rangle$.
Moreover the ground states can be labeled by chiral ring elements of $W$:
$$|\psi_j \rangle = \phi_j(z) |0\rangle$$
where the chiral ring is given by the monomials of $z$ modulo setting $\partial W=0$:
$$\phi_j(z)\in {\cal R} ;\qquad {\cal R}= {{\bf C}[z]\over \partial W}$$
To compute this, it turns out to be useful to introduce a basis of states known as the D-brane states \cite{Hori:2000ck}, which locally do not depend on the
parameters and so do not vary as we change parameters:
$$\langle \psi_i |\psi_j\rangle =\sum_{\alpha}\langle  \psi_i |D^+_\alpha\rangle  \langle D^-_\alpha |\psi_j\rangle$$
$ D^{\pm}_\alpha$ are identifed with lines in $z$ plane which when
projected to $W$-plane using $W(z)$, correspond to straight lines in $W$ plane emanating from the critical point and going to $\rm{Re}(W)=\pm \infty$.
In particular $\alpha$ labels the critical points.  
Then it can be shown that in the asymmetric limit where we rescale ${\overline W}\rightarrow \beta {\overline W} $ and set
$\beta \rightarrow 0$, we obtain \cite{Hori:2000ck}
$$\langle D^-_\alpha |\psi_j \rangle =\int_{D^-_{\alpha}} dz \ \phi_j(z) \ e^{W(z)}$$
and we find
$$\langle \psi_i |\psi_j\rangle =\sum_\alpha \big[ \int_{D^+_{\alpha}} d{\overline z} \ \phi_i({\overline z}) \ e^{-{\overline W(z)}}\big]\big[ \int_{D^-_{\alpha}} dz \ \phi_j(z) \ e^{W(z)}\big]=
\int d^2z \ \phi_i(\overline z) \phi_j(z) e^{W(z)-{\overline W}(\overline z)}$$
Where in the last equality we used the Riemann bilinear identity, which is somewhat of a formal step due to oscillatory nature
of the integral.  This result suggests that we can pretend {\it as if} the wave functions
are holomorphic and given by $\psi_j(z) \sim \phi_j(z) e^{W(z)}$, except that the complex conjugate wave function is not given by
the usual $\phi_j({\overline z}) e^{{\overline W}({\overline z})}$ but by $\phi_j({\overline z}) e^{-{\overline W}({\overline z})}$.  The oscillatory
nature of the integral which makes it convergent is precisely due to this change in sign of ${\overline W}$ making the exponent purely
imaginary.  We wish to emphasize that this is just an approximation to the actual wave function which has the usual definition
of inner product one is familiar with in the context of quantum mechanics.  One may ask in which limit is the computation of the
inner product exact? 
Each one is exact if we set the other $W$ to be small.  So in the limit that we rescale $W\rightarrow \beta  W $ and ${\overline W}\rightarrow \beta {\overline W}$
and send $\beta\rightarrow 0$ this inner product becomes exact\footnote{This is the UV limit of the theory in the context of 2d versions of this theory.},
 and in particular when $W$ is quasi-homogeneous
it is exact!
We can view the $D^+_\alpha$ as the analog of `conformal blocks' for this theory.  If we change the parameters of $W$ and bring it
back to itself, then the individual $D^+_\alpha$ undergo a transformation to a linear combination \cite{Hori:2000ck} because as the D-branes
defining the $D^+_\alpha$ cross one another, we get a Stokes phenomenon.  So when we come back to the original position
the $D^+_\alpha$ transform to a linear combination of the ones we had, and this gives a monodromy matrix\footnote{
In the language of the supersymmetric quantum mechanics, the eigenvlaues of the specific mondromy associated
with $W\rightarrow e^{2\pi i}W$ are given by $exp(2\pi iQ_R)$ where $Q_R$ are
the R-charges of the Ramond ground state.}:
$$D^+_\alpha \rightarrow M_\alpha^{\beta} D^+_\beta .$$
So the eigenvalues of the Berry's connection around loops can be computed exactly in this limit.
One may ask if there is any notion of $tt^*$ connection which is independent of taking any limits.
It turns out that $tt^*$ geometry has an `improved connection'  \cite{Cecotti:1991me}  
$$\nabla_i=D_i +C_i$$
where $D_i$ is the Berry's connection and $C_i$ is given by the action of $\phi_i$ on the vacua (for a recent
review and extension of $tt^*$ geometry see  \cite{Cecotti:2013mba}).  Unlike the Berry's connection, $\nabla$ is
flat for all parameters.  Since it does not depend on any parameters the monodromy of this improved connection can be computed
in this limit, which yields the above monodromy of the chiral wave functions.

\subsection{The Hamiltonian}

Now we come to the case of interest for us, and ask which Hamiltonian will give us the
wave function associated to the Liouville theory, which is motivated from its connection with
string theory, discussed in the next section.  This Hamiltonian has indeed been studied \cite{Cecotti:2014wea}.  Consider
$N$ quasi-particles each of which has in addition 4 degrees of freedom given by pairs of fermionic creation/annihilation operators
$b^{i\dagger}_{L,R},b^i_{L,R}$ where the only non-vanishing anti-commutators are
$$\{b^{i\dagger}_{L},b^j_{L} \}=\delta^{ij},\qquad \{b^{i\dagger}_{R},b^j_{R} \}=\delta^{ij} $$
and consider the Hamiltonian given by
$$H={1\over 2}\sum_i \big[ {\bf p}_i^2 + {\cal A}^i_{z}{\cal A}^i_{\overline z}\big]+{1\over \nu} \sum_{\substack{i,j\\  {i\not =j}}}\big[{b^{i\dagger}_L b^j_R\over (z_i -z_j)^2} +{ b^{j\dagger}_R b^{i}_L \over ({\overline z}_i -{\overline z_j})^2}\big] $$
where
$${\cal A}^i_{z}={1\over \nu}\sum_{\substack{j\\ j\not= i}} {1\over z_i-z_j},\qquad {\cal A}^i_{{\overline z}}={1\over \nu}\sum_{\substack{j\\ j\not= i}}{1\over {\overline z}_i-{\overline z}_j}$$
which corresponds to taking $W={1\over \nu} \sum_{i<j} {\rm log} (z_i -z_j)$.  
Thus, we have a Hamiltonian involving two particle and three particle interactions\footnote{Note that this Hamiltonian preserves Fermion number.  Moreover
one can show that the ground state has fermion number zero.  Thus if one is interested only in the ground state, one can restrict
the Hilbert space to the $2^N\subset 2^{2N}$ dimensional subspace.  Interestingly enough, this is the same as the dimension of a Hilbert
space for $N$ electrons, each of which has 2 spin states.}.   Moreover we can introduce quasi-holes at $\zeta_a$ by adding to $W$
$$W\rightarrow W+\sum_{i,a}{{\rm{log}}(z_i -\zeta_a)}$$
 The reason for labeling the terms ${\cal A}^i_{z}$ by ${\cal A}$, usually reserved for gauge potential,
is that if we consider $\overline{\partial_i} {\cal A}^i_{z}$ we get
$${\overline \partial_i} {\cal A}^i_{z}={2\pi i \over \nu}\sum_{\substack{j\\ {j\not= i}}} \delta (z_i -z_j).$$
This looks like a magnetic field and suggests that each particle has trapped $1/\nu$ units of magnetic flux.  This is somewhat
reminiscent of composite fermions model \cite{Jain:1989tx}. 

   However, this interpretation cannot be precise, because
the form of the interactions above is not the usual form one expects for electromagnetic interactions:
First of all, it does not contain ${\bf p}\cdot {\cal A}$ terms.  Moreover with the above definition of the gauge fields, the Field strength
is actually zero, because $\partial_z {\cal A}_{\overline z}=\partial_{\overline z} {\cal A}_z$.  One may be tempted to `gauge away' the
$|{\cal A}|^2$ term by redefining the wave function, but then this introduces ${\bf p}.{\cal A}$ terms. 
Nevertheless it can be written in a form, familiar in the context of Dirac operator, which behaves like a magnetic field.  Namely
we can rewrite $H$ as
$$H=Q^2$$
where
$$Q=\big[(b^{i\dagger}_L {\bf p}^i_z+b^i_L {\bf p}^i_{\overline z})+(b^i_R {\cal A}^i_{z}+b^{i\dagger}_R{\cal A}^i_{\overline z})\big]$$
If one can justify why the above Hamiltonian is a good description of fractional quantum Hall effect (for rational
values of $\nu$ and more specifically for $\nu ={n\over 2n\pm 1}$) in the context of superconducting boundary
conditions, we will have arrived at the conclusion of the previous section.
In particular,
this Hamiltonian, leads in the approximate sense we mentioned above, to the Liouville wave functions where
the corresponding blocks are given by computing
$$B_\alpha (\zeta_a, \nu )=\int_{D^-_\alpha} \prod_i dz_i \ {\rm exp} (W)=\int_{D^-_\alpha} \prod_i dz_i \ \prod_{i,a} (z_i -\zeta_a) \prod_{i<j} (z_i -z_j)^{1/\nu} $$
These are known to compute the conformal blocks of Liouville theory (see \cite{Dijkgraaf:2009pc,Cheng:2010yw}).  This in particular will undergo monodromy
as is expected for Liouville conformal theory.  For the 4-point quasi-holes we will get a 2 dimensional space, whose braiding
eigenvalues we already discussed for the 2 channels of the fusion of $\Phi_{1,2} \times \Phi_{1,2}$.  See in particular \cite{Gaiotto:2011nm} for
the explicit computation of this monodromy for the corresponding 4-pt function.\footnote{One may ask what is an effective Hamiltonian
describing the dynamics of the quasi-holes?  This gets related to the open string-wave function \cite{Aganagic:2011mi}.  It is argued in
\cite{Nekrasov:2014yra} that
this effective dynamics is captured by the Gaudin Hamiltonian (see also related discussions in \cite{Gaiotto:2011nm,Cecotti:2014wea}).}

\subsection{Adding back the magnetic flux and connections with the standard approach to FQHE}
The wave functions we obtain, using the Hamiltonians discussed above is very similar to the Laughlin type wave functions considered for FQHE, with one major difference:  In those cases one has in addition a non-holomoprhic term in the wave function given by
$$f(z_i)\ {\rm{exp}}(-B\sum_i |z_i|^2)$$
It is imperative for us that the wave functions be holomorphic, or at least meromorphic.  How
can we incorporate this in our setup?  The most natural thing to do in our context is to consider, instead of a uniform $B$ field, a lattice of fundamental units of $B$ fluxes at lattice points $\zeta_{ab}$.  In the usual description of the wave function this would have the effect of introducing 
$$\prod_{i,ab} {1\over (z_i-\zeta_{ab})}$$
This is reflected in our set up as follows:  The Liouville theory has a conservation law for momenta at zero chemical potential, which is violated by the curvature term by $(2g-2)(2\nu -1)$.   For simplicity
let us consider the $g=1$ case, so we will not have to worry about this, i.e. consider the theory on the torus.
In this case, putting $N$ fermions in the theory will violate the charge by $N$ units.  To cancel this in the usual Liouville context one puts a charge at infinity.  However, another way of doing this is to introduce
fluxes localized at points which is accomplished by introducing $exp[-i \sqrt{\nu} \phi(\zeta_{ab})]$ .  Let $\Phi$ denote the total number of such flux quanta.  Then Liouville conservation demands that
$$\nu \Phi= N.$$
This is the analog of the statement in the context of FQHE that for a given $\nu$, we have $N/\Phi =\nu$.

In the presence of such a term each particle picks up an additional term for $W$ given by
$$\delta W= -\sum {\rm log}(z-\zeta_{ab})$$
and to find the allowed ground state configuration we need to study the critical points 
$${d\delta W\over dz}=-\sum_{ab}{1\over (z-\zeta_{ab})}=0$$
which leads to $\Phi -1$ distinct critical points.  We need to distribute the $N$ states among these ground state
choices.  Since there are $N=\nu \Phi$ such particles even if we put fermionic statistics for them,
we would have ${\Phi-1\choose \nu \Phi}$ ground states for the Hamiltonian, which is a large number of states.  This is the analog of the problem which is faced in the standard approach to the FQHE where
there are $\Phi$ lowest Landau levels, and one needs to fill only $\nu \Phi$ of them.  In the
context of FQHE the question become which combination of these hugely degenerate states has the lowest energy, when one includes the electric repulsion between electrons.  In the context of the supersymmetric Hamiltonian we have been discussing, the ground states are degenerate and this can be viewed as the analog of turning off the electric repulsion.  Then the question in the present context becomes, which states among these hugely degenerate ground states of the supersymmetric system is `picked' when we turn on interactions.
It turns out that {\it there is} a distinguished state among the ground states of the supersymmetric Hamiltonian, and that is related to the fact that the operator state correspondence in this case maps the identity operator to a canonical ground state which is represented by the approximate holomorphic wave function of the Laughlin type we have already discussed.  That this is the one which will have the lowest energy when repulsion is included is natural because the identity operator is the combination of the critical points of $W$ where we take as spread out a combination of vacua as possible.  {\it 
It is interesting that in the present context the degeneracy of the lowest Landau level, is mapped to the degeneracy of the ground states of a supersymmetric system.} It would be interesting to see if this can lead to insights into FQHE based on supersymmetry.

In the presence of the magnetic fields the monodromy properties of the defects becomes more complicated as the quasi-holes will have to also go around $\zeta_{ab}$.  It would be interesting to work out the consequences of this.  However, it is clear that in the present context if we create a region in the sample
where $B$-flux is excluded, i.e. the $\zeta_{ab}$ are not placed in this region, the monodromy properties
of the quasi-holes we have discussed does not get modified and leads to that of the minimal model
monodromies.

\section{Connections with string theory and identifying the bulk theory}

So far I have tried to motivate the discussion from the viewpoint of the FQHE.   However,
to make the proposal more precise and identify the bulk theory, I need to explain the main motivation
for the present work. 
In section 4.1 I discuss the embedding in M-theory.  This is somewhat technical and is discussed
mainly to explain the motivation.  Readers
not familiar with string theory may wish to skip to section 4.2, which is largely independent,  where I spell out the proposal for the bulk theory.

This work arose from the realization of an unexpected similarity between what
one does in a special context in string theory and the structures involving FQHE.\footnote{These
ideas, and in particular the connection between FQHE and Liouville theory arising in Gaiotto theory were originally developed in discussions with Mina Aganagic and Sergei Gukov \cite{MinSer} to
whom I am grateful.}
It arose from the approach in \cite{Dijkgraaf:2009pc} (building upon the earlier work \cite{Dijkgraaf:2002fc,Dijkgraaf:2002dh})
 which relates the computation of supersymmetric
amplitudes in these theories (the `refined topological string amplitudes'), to $ADE$ matrix models and to chiral blocks of the corresponding Toda theory via the relation of matrix model to Toda theories.  In particular the topological string amplitudes (which can be viewed as wave functions \cite{bcov,witten}) correspond to the chiral blocks of
Toda theory, from which the supersymmetric amplitudes \cite{nekrasov,pestun} are obtained by taking their squares and integrating them.
The more natural  limit from the viewpoint of topological string, unlike the geometric case requiring $b^2>0$, is also $b^2 <0$ (as
we have been discussing in this paper) and $b^2=-1$ corresponds to the unrefined topological string.
Matrix amplitudes are of the same form as the holomorphic part of the Laughlin wave function, which is what originally
attracted my attention to a possible connection with FQHE.\footnote{The potential connection between matrix models and Laughlin wave functions
was pointed out to me by Shahin Sheikh-Jabbari in 2003.}

\subsection{Embedding FQHE in M-theory}

There has been many attempts to connect FQHE to modern developments in string theory \cite{Susskind:2001fb,Hellerman:2001yv,Bergman:2010gm,Fujita:2009kw,KeskiVakkuri:2008eb,Gaiotto:2008ak,Ganor:2014wua}.  
Here we propose a connection between supersymmetric ${\cal N}=2$ gauge theories
in 4-dimensions which arise from compactifications of 6-dimensional $(2,0)$ theories on a 2d Riemann surface (the `Gaiotto curve') \cite{Gaiotto:2009we} and FQHE.   

The $(2,0)$ theories are labeled by picking an ADE group.  Moreover we need to consider
the partition function of this theory on $S^3_{k,b^2}$,  where $b^2$ (sometimes written as $b^2=\epsilon_2/\epsilon_1$) denotes the squashing parameter of $S^3_k$,
and $S^3_k$ is the lens space $L(k,1)={S^3/ {\bf Z}_k}$.  In other words the worldvolume of the $6d \ (2,0)$ theories is taken to be
$${\rm ADE}\ (2,0) \ {\rm theory}\qquad on\qquad S^3_{k,b^2}\times {\bf R} \times \Sigma$$
The spacetime worldvolume of the FQH system is identified with ${\bf R}\times \Sigma$ where ${\bf R}$ is taken
as time and $\Sigma$ as the space.
We connect the rank of the corresponding $ADE$
with the number of layers for FQHE. 
 So the single layer case corresponds to the $A_1$ theory.  
Moreover the filling fraction is given by $\nu=-2\epsilon_2/\epsilon_1$.   We consider the Heegard decomposition of
$S^3_k$ to two solid tori (see \cite{Ooguri:1999bv,Witten:2010cx,Cecotti:2010fi,Dimofte:2010tz,Dimofte:2011jd} for the discussion
related to the present context) with suitable identification of boundary of solid tori depending on $k$.   Moreover $\epsilon_1,\epsilon_2$ can be viewed as the radii of the two circles of the middle torus.  There are two natural circles
in this geometry corresponding to the center of the two solid tori.  Note that each of these circles is associated
to one of the $\epsilon_i$ circles which does not shrink at the center of the tori, and we denote them by $S^1_{\epsilon_i}$.  Moreover this theory enjoys surface
operators which is fixed by choosing a point $\zeta_a\in \Sigma$ and a 2d plane of $S^3_{k,b^2}\times {\bf R}$, taken to be $S^1_{\epsilon_i}\times {\bf R}\subset S^3_{k,b^2}\times {\bf R}$.
From the view point of $\Sigma \times {\bf R}$ they can be viewed as two distinct types of defects located at $\zeta_a\in \Sigma$.
Picking the defects to be given by  $(r-1,s-1)$ copies of these two defects leads to the quasi-hole operators in the FQHE that
we discussed, namely $\Phi_{r,s}$. The conjecture by \cite{Alday:2009aq}\footnote{More precisely these
conjectures refer to $S^4$ geometry, but one can connect it to $S^3\times {\bf R}$ geometry
along the lines suggested in \cite{Nekrasov:2010ka}.} and extensions by \cite{Wyllard:2009hg}, adapted to this geometry \cite{Cordova:2013cea}, propose
that the partition function of these theories for $k=1$ case are given by Liouville theory and more generally by
$W_{ADE}$ Toda theories for the more general case which we identify with the `superconduting boundary condition' in FQH systems.  There are by now various derivations of these results 
\cite{Dijkgraaf:2009pc,Nekrasov:2010ka,Cordova:2013cea,Aganagic:2014kja}.
More precisely the effective theory on $\Sigma \times {\bf R}$ is expected to be the complexified version of the ADE Chern-Simons
theory with complexified level $k+is$, where $k$ is a positive integer
(labelling the level of $SU(2)\subset SL(2,{\bf C})$).  As was pointed out in \cite{witcs} unitarity is consistent with $s$ being either a real or purely imaginary number.  
We will be interested when $s$ is purely imaginary and write it as $s=-i\sigma$ with $\sigma >0$.  Interestingly enough the pure imaginary
case was also the main interest in \cite{witcs}.   The Liouville theory arises for $k=1$, i.e. when we have $S^3$.   The  Liouville theory on boundary
of the space (with signature 1+1, which can be interpreted as edge modes) arises in this construction by a specific boundary condition for fields \cite{Gaiotto:2011nm}.   For a discussion of $A_1$ case
see \cite{Verlinde:1989ua,Teschner:2005bz} in the context of real $SL(2,R)$ Chern-Simons theory and \cite{Dimofte:2011jd,Cordova:2013cea}
for the complex case.  See also the discussion in the next section.

Our general setup
naturally allows for the lens space extension given by $k\geq 1$ and recently studied in \cite{Dimofte:2014zga,Gukov:2015sna}.
For more
general $k$ the corresponding conformal theory we get is roughly a mixture between Liouville theory and an extra parafermionic system which enjoys
a $Z_k$ symmetry.  For general $k$ the $b^2$ of Liouville is related to it by \cite{Dimofte:2010tz,Dimofte:2014zga,Gukov:2015sna}\footnote{I thank T. Dimofte and S. Gukov for discussions on this.}
$${2\over k-\sigma}={1+b^2\over k},\qquad {2\over k+\sigma}={1+b^{-2}\over k}.$$ 

\subsection{Complex Chern-Simons Theory as the effective theory of FQHE}
The discussions up to now can be viewed as motivations for the statement to be
made in this section, as we collect the various observations made in the previous sections to make our proposal
precise.  It is well known that FQH system in the IR can be described by a topological
theory, Chern-Simons theories being the prototypical examples.  We propose that:

{\it The effective IR theory describing a single layer FQH system is Chern-Simons theory
based on} $SL(2,{\bf C})$ {\it (as well as its supersymmetric version)}.

More precisely we propose that the prinicpal series with filling fraction 
$\nu =n/(2n\pm 1)$
(i.e. $b^2=-(2n\pm 1)/2n$) are
described by complex Chern-Simons theory $SL(2,{\bf C})$ with
$$(k,\sigma)=(\pm 1,4n\pm1)$$
$$\qquad {k+\sigma\over 2}=2n\pm1, \qquad -{k-\sigma\over 2} =2n$$
(and similarly for supersymmetric version leading to the filling fractions $\nu= m/ (m+2)$ which we leave
to the interested reader).
The action for the CS theory is given by \cite{witcs}
$$S={(k-\sigma) \over 8\pi}\int_M {\rm Tr} \big( {\cal A}d{\cal A}+{2\over 3} {\cal A}^3\big)+{(k+\sigma)\over 8 \pi}\int_M {\rm Tr} \big({{\cal \bar A}}d{\bar {\cal A}}+{2\over 3} {\bar {\cal A}}^3\big)$$
$$={-2n\over 4\pi}\int_M {\rm Tr} \big( {\cal A}d{\cal A}+{2\over 3} {\cal A}^3\big)+{(2n\pm1)\over 4\pi}\int_M {\rm Tr} \big({{\cal \bar A}}d{\bar {\cal A}}+{2\over 3} {\bar {\cal A}}^3\big)$$
where ${\cal A}$ is a complex $SL(2,{\bf C})$ connection.  The action almost splits, between
${\cal A}$ and $\bar {\cal A}$.  However, the fact that one is complex conjugate of the
other is what couples them in a non-trivial way.  This is the formulation of the theory
in Euclidean three dimensional space.  In the physical context of the $2+1$ signature,
instead of $SL(2,{\bf C})$ we have $SL(2,{\bf R})\times SL(2,{\bf R})$, where the
two connections can be written as
$${\cal A}=\omega - e \qquad {\bar {\cal A}} =\omega +e$$
and $(\omega, e)$ are independent, but on-shell they get related and interpreted in terms of the spin connection and vierbein respectively in the gravitational context.
The coefficient $k$ of Chern-Simons terms has to
be integer:  In that case we have $\omega$ identified with $SO(2)\subset SL(2,{\bf R})_{diagonal}$
and integrality of spin demands quantization of $k$.  In gravitational context $e$ is not quantized.
     In particular $\sigma$ (related to cosmological constant) is not quantized.  Nevertheless it is amusing
that we are finding that the quantized values of $\sigma$ which would have been 
natural if the two $SL(2,R)$'s have independently integral $SO(2)$ charges leads {\it precisely} to the realized cases in FQHE!
Changing $\sigma$ in the context of FQHE can be viewed as changing the density of electrons (or equivalently
the $B$-field) and so in a sense
arbitrary values are also natural from the perspective of FQH applicationl.  
 The fact that the levels differ by one unit
simply reflects the fact that we have identified the principal series with $|k|=1$, and
we can of course consider other relative shifts between the two levels.

 In FQHE chiral modes correspond to gapless edge modes (see \cite{Wen} for a discussion and references), thus it is natural to consider this theory in the presence of a boundary.  This will lead to different CFT's on the boundary with different
values of $(c_L,c_R)$.    One boundary condition \cite{DGV}, which  we identify with sharp edge in the FQH system,
corresponding to the polarization one usually chooses in quantization of it \cite{witcs} and given by
$${\delta \Psi \over \delta A_{\bar z}}=0={\delta \Psi \over \delta {\bar A}_z},$$
gives
$$(c_L,c_R)=\big( 3-{12\over k+\sigma}, 3+{12\over k-\sigma}\big)=\big( 3-{6\over 2n},3-{6\over {2n\pm 1}}\big)$$
The usual $(c_L,c_R)$ for a compact group would have been \cite{Witten:1988hf}
$$(c_L,c_R)=\big({k\ {\rm dim} G\over k+h},0\big)$$
Note that this is a rather unusual situation where {\it the chiral blocks do not factorize} between
left-moving and right-moving sector.  This phenomenon was discovered in the beautiful paper
\cite{witcs} and is a key point for our model of FQHE. 
This boundary condition we will identify with a sharp edge of the FQH system, which
can support edge currents.  In quantizing the Hilbert space there is one left- and one
right-moving $U(1)_{L,R}$  which
are the Cartan of the two $SL(2,{\bf R})$'s.  We identify the current of $U(1)_R$ with electric current.
The reason for this is that as was shown in eq.(5.3) of \cite{witcs}  (see also 
\cite{Axelrod:1989xt}) as far as the dependence of the
partition function on the $U(1)$ charges, the characters will include $\theta (\tau,z; \bar \tau ,\bar z)$ functions
of level $(2n, 2n\pm1)$ relative to $\tau$ and $\bar \tau$.  This implies that
$U(1)_R$ will have fractions ${r\over {2n\pm 1}}$ which is consistent with our description of the
system as realizing FQHE with $\nu ={n\over 2n\pm 1}$.    The fact
that in the bulk they should reproduce the non-trivial braidings of $SL(2,{\bf C})$ shows that they realize minimal model operator algebra, leading
to non-abelian braidings that we have already mentioned.  In particular as we take the basic quasi-hole around another
one, we get two possible channels\footnote{Except possibly for the $\nu=1/3$, as discussed below.} instead of one channel expected in abelian models, with phases 
$$\big( e^{2\pi i n\over 2n\pm1}, e^{-{6\pi in\over 2n\pm1}} \big)$$
Let us consider $(c_L,c_R)$ for the first few cases:  
$$\nu={1\over 3}:\qquad (0,1)$$
$$\nu={2\over 3}:\qquad ({3\over 2},1)$$
$$\nu={2\over 5}:\qquad ({3\over 2},{9\over 5})$$
$$\nu={3\over 5}:\qquad (2,{9\over 5})$$
$$\ldots $$
$$\nu\rightarrow {1\over 2}:\qquad (3,3)$$
Naively one would expect that the unitarity completely fixes the first one to be the usual $c=1$ and with the radius of the boson fixed
by the level of $\theta $ function to be the usual description of Laughlin's system of a chiral
boson at radius $R^2=1/3$.  However, even this case needs to be further studied:  As pointed out in \cite{witcs}
for pure imaginary $s$, which is the case of interest for us, there is an `exotic hermitian structure' on the Hilbert space.  Nevertheless
it was shown there, that at least in genus 1 case, which is the only case of relevance for our consideration in the $1+1$ dimensional
edge theory,
there is in addition a {\it unitary structure} if  $0<|{\sigma-k\over \sigma +k} |^{sgn(k)}<1$ and for our case this is given by
$$ 0<\big| {\sigma -k\over \sigma+k} \big|^{sgn(k)}=\big( {2n\over 2n\pm1}\big)^{\pm 1}=(2\nu_\pm)^{\pm 1}<1$$
which is nicely satisfied.  It is interesting that unitarity is related to properties of filling
fraction for the principal series.  It is clear that
the unitarity structure for the problem at hand is rather special  and one cannot simply borrow the technology
familiar from the compact WZW models to the case at hand.  In addition, familiar corrections such as
$k\rightarrow k+h$ in the compact case, do not apply to the non-compact case \cite{witcs}.
Moreover our model has, in addition to quasiholes, a rich structure of other modes with continuous wave-like modes in these blocks.
 We leave a detailed study of this non-trivial block structure and the corresponding character to a future work \cite{DGV}.  Regardless
of the explicit structure of the left-right mixed blocks,
it is true that the braiding properties of the defects are all determined by the
bulk $SL(2,{\bf C})$ theory and as we have discussed, as a part of that, the quasi-hole defects realize the
 operator algebra for $(2n,2n\pm1)$ minimal models  and the associated braiding structure.

One can ask why are the results we are obtaining so different from the conventional picture of
the principal series?  A hint comes from the fact that for $\nu =1/3$ our description seems
to be particularly close to the usual one.  In paricular, the fact that we are getting a system with $(c_L,c_R)=(0,1)$
suggests that the wave function for the electrons are chiral and given by the original
Laughlin wave-function which is holomorphic\footnote{The same is true for the Moore-Read state, where our construction
would yield a $c_L=0$.}.  However, for higher values of $(c_L,c_R)$, a key feature of our model
is that blocks are {\it not} purely holomorphic or anti-holomorphic.  Therefore  we expect that for higher $(c_L,c_R)$ the electron wave function is {\it not purely holomorphic} and it should also have an anti-holomorphic dependence.  We are currently studying details of this for our model.
%\footnote{The operator $\Phi_{2n+1,{2n\pm 1}+1}$,
%which is just beyond the Kac-table (whose interior characterizes quasi-hole states), behaves in many ways like
%the operator corresponding to the electron.}
  We thus believe that the assumption of holomorphic
projection of the wave function, which is usually assumed in the conventional approach, is what sets it apart from ours, except for the $\nu={1/ 3}$ case.  It would be interesting to revisit the validity of this assumption in solving for the electron wave function in numerical analysis of FQH systems for other
values of $\nu$.

There are other boundary conditions possible for this theory.
In general, different boundary conditions can affect $(c_L,c_R)\rightarrow (c_L-c,c_R-c)$ by some $c$
which masses up some left-right chiral modes.  Note that $c_L-c_R$ is invariant (at least mod $24$ due to gravitational anomaly \cite{witcs}), and from the
above formulas we obtain
$$c_L-c_R= {6Q^2\over k}$$
where $Q^2=(b+{1\over b})^2$.  Let us specialize to $k=1$ case.  In that case we get
$$c_L-c_R=c_{Liouville}-1$$
because $c_{Liouville}= 1+6Q^2$.  So this suggests that we should be able to get Liouville theory
by a suitable boundary condition, and indeed this is the case \cite{Gaiotto:2011nm,Cordova:2013cea} . Moreover, it also shows that there is a left-over right-moving piece with $c_R=1$,
which ends up being simply a free right-moving boson \cite{Cordova:2013cea}\footnote{If
the boundary of Riemann surface is an annulus, which corresponds to the $S^4$ partition function \cite{Nekrasov:2010ka},
then we will have both a left-moving and a right-moving Liouville as well as a left-moving and a right-moving boson.
We identify this as placing the FQH system in contact with two superconducting boundary components.
This makes contact with AGT.
Indeed it was already noted \cite{Alday:2009aq} that in addition to Liouville they needed an extra $U(1)$ boson to
get the correct partition function of 2 M5 branes on $S^4$, consistent with the extra $c=1$ system.}.  Note that with this boundary condition, there is no interesting current.
The reason for this identification is that the boundary condition leading up to Liouville theory, sets the upper-triangular
part of $SL(2,{\bf C})$ current $\langle J^+\rangle \not=0$ which breaks the $U(1)$ symmetries, i.e. higgses the $U(1)$'s, one of which we are identifying
with the EM current.  This can only happen if the EM is itself in a superconducting phase.
We thus identify this boundary condition as placing the material next to a superconductor.
 We therefore predict that in these cases we should have a rich structure of neutral currents (for a discussion
of placing FQH material next to a superconductor see e.g.\cite{Alicea:2014lta}). There should be as many channels
as the blocks of the minimal model $(2n,2n\pm 1)$, without any electric currents.

This brings us full circle back to our original motivation:  We connected
the choices of filling fractions for principal series to the unitary minimal models
of Virasoro and superVirasoro.  The connection here is that the choice of boundary condition
which this entails gets rid of right-moving electric modes and leads to a chiral theory with $c<1$,
which can be classified using unitarity.   With the sharp edge boundary we get, as already mentioned,
a related theory with $(c_L,c_R)> (1,1)$ (except for the very first one, which is $(0,1)$).  It is then natural that with the sharp edge boundary
we get filling fractions which are compatible with the unitarity of  superconducting edge theory {\it in case we had put the sample
next to a superconductor!}

In the sharp edge boundary, the electric current can be
carried only in one direction in our model.  The electric current is carried by the $U(1)_R$
(called the `downstream') and the left-moving direction (the `upstream'), is where the neutral
mode propagates coming from our $SL(2,{\bf R})_L$ modes.  Our model therefore predicts that the electric current can {\it only} go in one direction
along the edge (correlated with the magnetic field).
It turns out that standard models that were originally constructed had electric currents going
in both directions (for half of the principal series).  These were not found in experiments.  Later it was found 
how to remedy this and obtain electric currents which only go downstream, at least in some cases, by going to a new phase by taking into account disorder \cite{Kane:1994vb}.
It is a nice feature of our model that it {\it predicts}\footnote{For predictions
of transverse response functions based on other models see in particular \cite{sasha}.} electric currents moving {\it only} downstream
 {\it and} neutral currents moving upstream in essentially all cases.  The proliferation of neutral upstream
currents is what our model predicts due to many left-moving
modes.  Recent experimental results
\cite{amir,weiz} confirm such upstream neutral currents which is difficult to reconcile with standard models
of FQHE for filling fractions $\nu={n\over 2n+1}$ which has no upstream currents.

Moreover for the standard series with filling fraction $\nu ={n\over 2n\pm 1}$ we have the
left- and right- moving central charges which will have their imprint on thermal Hall conductivity \cite{Kane:1997fda}, which (in fundamental units) is given by
$$\sigma_H = c_L-c_R= {\mp 6\over 2n(2n\pm 1)}$$
Note that $R$ refers to downstream direction throughout this paper.
Studying the thermal
transport properties of the edge currents
is another way to distinguish this model from other ones\footnote{We would like to thank
T. Dumitrescu for discussions on this.}.  Moreover the experimental results
in \cite{amir}, where one measures left- and right-moving powers
may be one way to experimentally measure $(c_L,c_R)$.

 In our model we have two levels $(k,\sigma)$, and the quantization
of $k$ is required for the theory to be well defined, but $\sigma$ is not, and changing
it corresponds to changing the electron density.  This allows us to move away
from rational filling fractions and consider the transitions between Plateau regions in FQHE,
which is a nice feature of our model, and can be potentially used
to study universality properties in quantum Hall transitions.  In fact the results of studies of $SL(2,{\bf C})$ Chern-Simons
knot invariants and their jumping phenomenon of the free energy as a function of $b^2$ (which is
related to inverse of the filling fraction) seems to be interpretable \cite{DGV} as the relation between
resistivity and the $B$-field in the FQHE (see in particular (fig. 4) of \cite{zag}, where the vertical
axis can be related to the B-field and the horizontal to the resistivity).

One can also study situations where the whole construction is lifted up one dimension.
In fact, in a sense this is forced on us if we wish to also change the value
of $k$ continuosly.
This could arise only in a context where the FQH system is a boundary of a 3d material where the
boundary theory cannot decouple from the 3 dimensional bulk theory at non-integral $k$, similar to what one encounters
in the context of topological insulators or 2d Dirac/Weyl metals.  Then we could ask what is the
effective theory in the 3+1dimensional bulk theory which couples to $SL(2,{\bf C})$ theory (or more generally
complex $ADE$ Chern-Simons theory) which fixes the issue related to non-integrality of $k$ on the 2+1 dimensional boundary?  This has been answered by Witten:
 The effective
theory in this context would involve ${\cal N}=4$ topologically twisted $ADE$ gauge symmetry in $3+1$ dimensions (in this case $A_1$), which can altenratively be viewed as $ADE$ complexified gauge theories in $3+1$ dimensions, whose boundary theory is known to give the corresponding 2+1 Chern-Simons theory
\cite{wit4,Witten:2011zz}.  This would allow us to move away from $k$ being an integer.  Note that even though the underlying
theory is supersymmetric, the topologically twisted version treats this as a BRST symmetry and the only
manifestation of supersymmetry is in its topological properties.
Moreover the choices of interesting 2+1 material which can be placed as an interface
in this system translates to the choices of consistent
boundary conditions for supersymmetric Yang-Mills theory, and this has been analyzed in \cite{gw},
leading to a rich structure.  A key role there is played by the Montonen-Olive $SL(2,{\bf Z})$ symmetry
\cite{Montonen:1977sn} of ${\cal N}=4$ supersymmetric Yang-Mills theory for non-abelian gauge theories which leads to duality symmetries on the boundary.\footnote{The abelian version of this at half-integral points of the Montonen-Olive torus is presumably related to dualities studied in the condensed matter literature.}  It is
interesting that our model of FQHE naturally suggests the {\it potential experimental realization of 
topologically twisted non-abelian Yang-Mills theories} as effective descriptions in the bulk of topological insulators!
We are currently studying potential applications of this to condensed matter systems \cite{dgv}, which
seems to also lead to beautiful block structures\footnote{In the simplest cases this is related to sections of suitable bundles on Hitchin moduli space \cite{gukw} on the boundary surface.} on the $2+1$ boundary, analogous to the $1+1$ edge
modes for the usual FQH systems.

It is natural to study $k>1$ systems as well and see whether they give interesting
examples of filling fractions observed in experiments.
For example filling fractions $\nu={n\over (2nk'\pm1)}$, known as the Jain series,
are anticipated in the composite fermions theory \cite{Jain:1989tx}.  It is natural to conjecture that the higher Jain series correspond in our setup to the two $SL(2,{\bf R})$ levels being given by $(-2n,2nk'\pm 1)$,
which would correspond to the level $k=2n(k'-1)\pm1$ for the diagonal $SL(2,R)$.
 In addition we can have more general values of $k$ which
would presumably give new series.  This would be worth investigating.    For $k>1$ the superconducting boundary
conditions are expected to lead to paraLiouville theories \cite{tach} (see \cite{NW} for a discussion
of para-Liouville theories).
 Similarly the supersymmetric case for $k\geq 1$ would
be interesting to study.

\subsection{Multi-layer case}
Let us briefly discuss the connection with Toda case.  Most of our discussion before was in the context of $A_1$ theory
which gives the single layer theory.\footnote{As an interesting example of theoretical study of a bi-layer system see \cite{Wang}.}
We will restrict our attention to the non-supersymmetric
case, though given what we found for the $A_1$ case it is expected that the supersymmetric case of Toda theories are also interesting.\footnote{
From the expressions below it is natural to guess that the filling fraction matrix is given by ${2m\over m+2} C^{-1}$ for the supersymmetric case.}
Our conjecture maps the multi-layer FQHE to $ADE$ Toda system, when we put the
sample in interface with a superconductor: 
$$\int_{\Sigma} d^2z\big[ {1\over 8\pi}\partial {\vec \phi}\cdot {\overline \partial}{\vec \phi}+i Q{\vec \rho}\cdot {\vec \phi}R+\sum_{j=1}^{r_{ADE}} {\rm exp}(b\ {\vec e}_j\cdot {\vec \phi})\big]$$
where $\vec \phi$ is an r-dimensional vector parameterizing the Cartan of $ADE$, $e_j$ are the simple roots, $Q=b+{1\over b}$, $\vec \rho$ is half the sum of positive roots of $ADE$ and the central charge  of the Liouville theory is given by $c=r+12 {\vec \rho}\cdot {\vec \rho}(b+{1\over b})^2$.  
 The correponding holomorphic blocks will involve terms of the form
$$\prod_{\substack{i,j,a,b}} (z_i^\alpha-z_j^\beta)^{-b^2 C_{\alpha\beta}}$$
where $\alpha,\beta$ label the layers, and is as many as the rank of $ADE$.  Moreover $C_{\alpha \beta}$ is the Cartan matrix.\footnote{To make the wave function
have no poles, we may have to choose a different basis for fields.}
 These theories possess $W_{ADE}$ symmetry
algebra (see \cite{Fateev:1987zh} for the $A_{N-1}$ case).  For simplicity let us focus on the $A_{N-1}$ case and specialize to the minmal model
case (which has been found to lead to interesting structure in the AGT setup in \cite{Estienne:2011qk,Bershtein:2014qma,Alkalaev:2014sma}):  This corresponds to $W_N(p,p+1)$ models which
map to $b^2=-{2p\pm 1\over 2p}$, with central charge 
$$c=(N-1)\big[ 1-{N(N+1)\over 2p(2p\pm1)}\big]$$
with $p\geq N$.  This has the chiral wave function
$$\prod_{\substack{i,j,a,b}} (z_i^\alpha-z_j^\beta)^{{2p\pm 1\over 2 p}C_{\alpha\beta}}$$

which leads to the filling fraction matrix $\nu$ given by
$$\nu ={2p\over 2p\pm 1}C^{-1}$$
with $2p\geq N$, where $N-1$ is the number of layers and $C^{-1}$ is the inverse of the Cartan
matrix for $A_{N-1}$.  The single layer case corresponds to $C=2$ which gives the filling fractions
of the principal series $\nu ={2p\over 2p\pm 1}$.  This formula for $\nu$ is valid for all ADE Toda cases where
$C$ is the corresponding Cartan matrix.
It would be interesting to connect this to experimental observations for multi-layer FQHE. 
Of course, just as in the case of the single layer, the sharp edge boundary condition
will give a different theory, which differs from chiral Toda theories, but which
will still have the same $c_L-c_R$.

\subsection{Punctures}
Connections with string theory suggest that there is more one can do.  In particular in the context of Gaiotto theories,
we are led to put regular and irregular punctures on the Riemann surface.   In the language of FQHE
these should be creating some `regular and irregular defects' in the sample!  It would be interesting to 
try and realize these.  In the single-layer case the regular defects are equivalent to excizing a point  $w$ from
the surface and considering the chiral wave function $\prod_i{1\over(z_i-w)^m}$.  The irregular ones
are most naturally placed at the boundary of the space and involve a boundary term $ {\rm exp}\big[\oint W(z) \partial \phi \big]$ \cite{Dijkgraaf:2009pc} where $W=z^n$.
It would be interesting to find realizations of these ideas in the FQHE context.  Moreover, the $k>1$ versions of these would be expected to also exist
as in the $A_1$ case.

\subsection{5d systems and anisotropic FQHE}
There is a 5d version of these supersymmetric theories which get connected to q-deformed versions of Toda theories.
In these cases from the results in  \cite{Aganagic:2012hs} one expects that the zeroes of the Laughlin wave function get split.
In particular in this case the holomorphic part of the wave function  for filling fraction $\nu =1/m$ instead of $(z_1-z_2)^m$ is given by
$$\prod_{i=1}^{m}\big[q^{i/2}{\rm exp}(x_1)-q^{-i/2}{\rm exp}(x_2)\big]$$
where $z_{1,2}={\rm exp}(x_{1,2})$ and $q$ is an additional deformation parameter.  It is conceivable that this is relevant for the anisotropic versions of FQHE.  Moreover from
the fact that $z_i$ are replaced by their logs, it is suggestive that these are related to cylindrical geometries for the FQHE.
These should have intersting physical realizations!

\vskip 0.5cm

\hspace*{-0.8cm} {\bf\large Acknowledgements}

\vskip 0.2cm
I would like to thank M. Aganagic and S. Gukov for participation at initial stages of this project.
I am grateful to A. Abanov and X.-G. Wen for extensive discussions on
various aspects of FQHE as well as to S. Cecotti and R. Dijkgraaf for discussions and various past
collaborations which were highly influential in my thinking about this subject.
I would also like to thank M. Cheng, T. Dimofte, T. Dumitrescu, M. Fisher, B. Halperin, D. Harlow, D. Jafferis, G. Moore, N. Read, S. Sachdev, A. Semikhatov, S. Shakirov, J. Teschner, E. Witten
and A. Yacoby for insightful comments. And,
last but not least, I would like to thank my son Farzan Vafa for discussions on FQHE which
rekindled my interest in this topic.

This research is supported in part by NSF grant PHY-1067976.

\end{document}